\documentclass[preprint,12pt]{elsarticle}
\usepackage{geometry}                
\usepackage{graphicx}
\usepackage{texnames}
\usepackage[T1]{fontenc}
\usepackage{mathrsfs}
\usepackage{amssymb}
\usepackage{amsmath}
\usepackage{amsfonts}
\usepackage{epstopdf}
\usepackage{subfigure}

\journal{Astroparticle physics}

\begin{document}

\begin{frontmatter}
\title{Estimating the significance of a signal in a multi-dimensional search}

\author{Ofer Vitells}
\ead{ofer.vitells@weizmann.ac.il}
\author{Eilam Gross}
\ead{eilam.gross@weizmann.ac.il}

\address{Weizmann Institute of Science, Rehovot 76100, Israel}

\begin{abstract}
In experiments that are aimed at detecting astrophysical sources
such as neutrino telescopes, one usually performs a search over a
continuous parameter space (e.g. the angular coordinates of the sky,
and possibly time), looking for the most significant deviation from
the background hypothesis. Such a procedure inherently involves a
``look elsewhere effect'', namely, the possibility for a signal-like
fluctuation to appear anywhere within the search range. Correctly
estimating the $p$-value of a given observation thus requires
repeated simulations of the entire search, a procedure that may be
prohibitively expansive in terms of CPU resources. Recent results
from the theory of random fields provide powerful tools which may be
used to alleviate this difficulty, in a wide range of applications.
We review those results and discuss their implementation, with a
detailed example applied for neutrino point source analysis in the
IceCube experiment.
\end{abstract}


\end{frontmatter}

\section{Introduction}
The statistical significance associated with the detection of a
signal source is most often reported in the form of a $p$-value,
that is, the probability under the background-only hypothesis of
observing a phenomenon as or even more `signal-like' than the one
observed by the experiment. In many simple situations, a $p$-value
can be calculated using asymptotic results such as those given by
Wilk's theorem \cite{wilks}, without the need of generating a large
number of pseudo-experiments. This is not the case however when the
procedure for detecting the source involves a search over some
range, for example, when one is trying to observe a hypothetic
signal from an astrophysical source that can be located at any
direction in the sky. Wilk's theorem does not apply in this
situation since the signal model contains parameters (i.e. the
signal location) which are not present under the null hypothesis.
Estimation of the $p$-value could be then performed by repeated
Monte Carlo simulations of the experiment's outcome under the
background-only hypothesis, but this approach could be highly time
consuming since for each of those simulations the entire search
procedure needs to be applied to the data, and to establish a
discovery claim at the $5\sigma$ level
($p$-value=$2.87\times10^{-7}$) the simulation needs to be repeated
at least $\mathcal{O}(10^7)$ times. Fortunately, recent advances in
the theory of random fields provide analytical tools that can be
used to address exactly such problems, in a wide range of
experimental settings. Such methods could be highly valuable for
experiments searching for signals over large parameter spaces, as
the reduction in necessary computation time can be dramatic. Random
field theoretic methods were first applied to the statistical
hypothesis testing problem in~\cite{davies}, for some special case
of a one dimensional problem. A practical implementation of this
result, aimed at the high-energy physics community, was made
in~\cite{epjc}. Similar results for some cases of multi-dimensional
problems~\cite{adlerhasofer}\cite{adler1} were applied to
statistical tests in the context of brain imaging~\cite{worsley}.
More recently, a generalized result dealing with random fields over
arbitrary Riemannian manifolds was obtained~\cite{adler}, openning
the door for a plethora of new possible applications. Here we
discuss the implementation of these results in the context of the
search for astrophysical sources, taking IceCube \cite{icecube} as a
specific example. In section~\ref{sec1} the general framework of an
hypothesis test is briefly presented with connection to random
fields. In section~\ref{sec2} the main theoretical result is
presented, and an example is treated in detail in
section~\ref{sec3}.

\section{Formalism of a search as a statistical test}\label{sec1}

The signal search procedure can be formulated as a hypothesis
testing problem in the following way. The null (background-only)
hypothesis $H_0:\mu=0$, is tested against a signal hypothesis
$H_1:\mu>0$, where $\mu$ represents the signal strength. Suppose
that $\theta$ are some nuisance parameters describing other
properties of the signal (such as location), which are therefore not
present under the null. Additional nuisance parameters, denoted by
$\theta'$, may be present under both hypotheses. Denote by
$\mathscr{L}(\mu,\theta,\theta')$ the likelihood function. One may
then construct the profile likelihood ratio test
statistic~\cite{asimov}

\begin{equation}\label{eq:q}
q = -2\log \frac{\displaystyle\max_{\theta'}
\mathscr{L}(\mu=0,\theta')}{\displaystyle\max_{\mu,\theta,\theta'}
 \mathscr{L}(\mu,\theta,\theta')}
\end{equation}

\noindent and reject the null hypothesis if the test statistic is
larger then some critical value. Note that when the signal strength
is set to zero the likelihood by definition does not depend on
$\theta$, and the test statistic (\ref{eq:q}) can therefore be
written as

\begin{equation}\label{eq:qtheta}
q =  \displaystyle\max_{\theta \in \mathscr{M}} q(\theta)
\end{equation}

\noindent where $q(\theta)$ is the profile likelihood ratio with the
signal nuisance parameters fixed to the point $\theta$, and we have
explicitely denoted by $\mathscr{M}$ the $D$-dimensional manifold to
which the parameters $\theta$ belong. Under the conditions of Wilks'
theorem~\cite{wilks}, for any fixed point $\theta$, $q(\theta)$
follows a $\chi^2$ distribution with one degree of freedom when the
null hypothesis is true. When viewed as a function over the manifold
$\mathscr{M}$, $q(\theta)$ is therefore a $\chi^2$ \emph{random
field}, namely a set of random variables that are continuously
mapped to the manifold $\mathscr{M}$. To quantify the significance
of a given observation in terms of a $p$-value, one is required to
calculate the probability of the maximum of the field to be above
some level, that is, the excursion probability of the field:

\begin{equation}\label{eq:pval}
p\text{-value}=\mathbb{P}[ \displaystyle\max_{\theta \in
\mathscr{M}} q(\theta)
> u].
\end{equation}

Estimation of excursion probabilities has been extensively studied
in the framework of random fields. Despite the seemingly difficult
nature of the problem, some surprisingly simple closed-form
expressions have been derived under general conditions, which allow
to estimate the excursion probability (\ref{eq:pval}) when the level
$u$ is large. Such `high' excursions are of course the main subject
of interest, since one is interested in estimating the $p$-value for
apparently significant (signal-like) fluctuations. We shall briefly
describe the main theoretical results in the following section. For
a comprehensive and precise definitions, the reader is referred to
Ref. \cite{adler}.

\section{The excursion sets of random fields}\label{sec2}

\begin{figure}[ht!]
\begin{center}
\includegraphics[width=12cm]{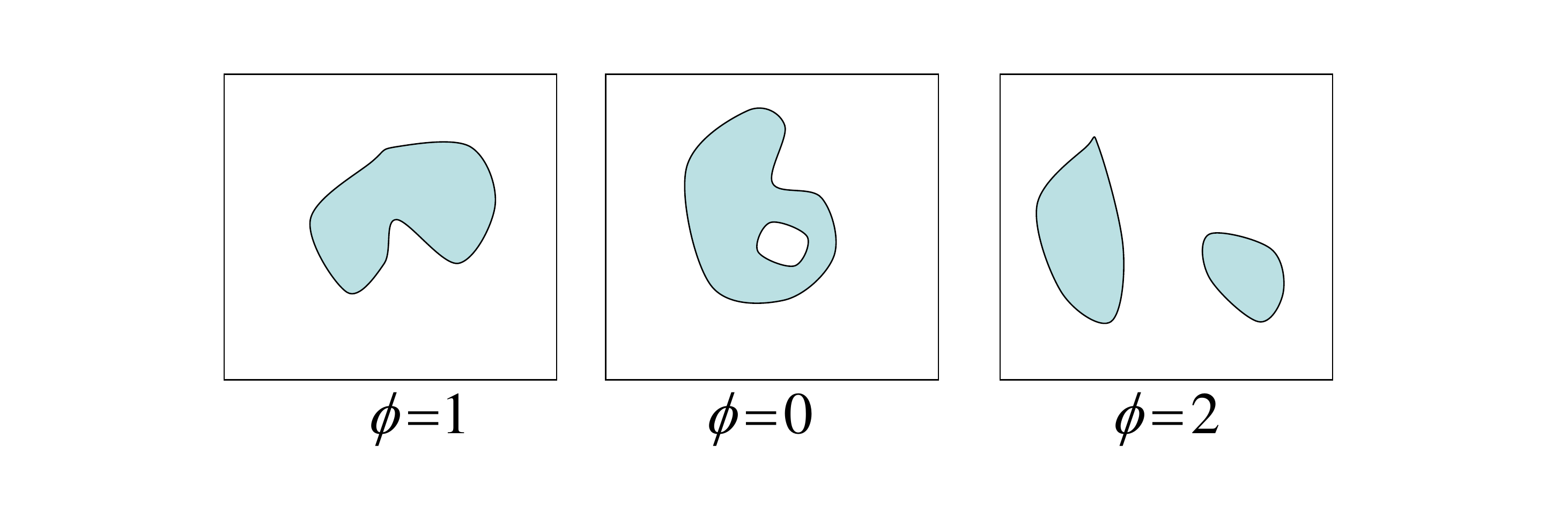}
\end{center}
\caption{Illustration of the Euler characteristic of some
2-dimensional bodies. }\label{fig:eulerillus}
\end{figure}

The excursion set of a field above a level $u$, denoted by $A_u$, is
defined as the set of points $\theta$ for which the value of the
field $q(\theta)$ is larger than $u$,

\begin{equation}\label{eq:Au}
A_u = \{\theta \in \mathscr{M} : q(\theta) > u \}
\end{equation}

\noindent and we will denote by $\phi(A_u)$ the \emph{Euler
characteristic} of the excursion set $A_u$. For a 2-dimensional
field, the Euler characteristic can be regarded as the number of
disconnected components minus the number of `holes', as is
illustrated in Fig.\ref{fig:eulerillus}. A fundamental result
of~\cite{adler} states that the expectation of the Euler
characteristic $\phi(A_u)$ is given by the following expression:

\begin{equation}\label{eq:euler}
\mathbb{E}[\phi(A_u)] = \sum_{d=0}^D \mathscr{N}_d \rho_d(u).
\end{equation}

The coefficients $\mathscr{N}_d$ are related to some geometrical
properties of the manifold and the covariance structure of the
field. For the purpose of the present analysis however they can be
regarded simply as a set of unknown constants. The functions
$\rho_d(u)$ are `universal' in the sense that they are determined
only by the distribution type of the field $q(\theta)$, and their
analytic expressions are known for a large class of `Gaussian
related' fields, such as $\chi^2$ with arbitrary degrees of freedom.
The zeroth order term of eq. (\ref{eq:euler}) is a special case for
which $\mathscr{N}_0$ and $\rho_0(u)$ are generally given by

\begin{equation}\label{eq:zero}
\mathscr{N}_0 = \phi(\mathscr{M}), \hspace{0.5cm} \rho_0(u) =
\mathbb{P}[q(\theta) > u]
\end{equation}

\noindent Namely, $\mathscr{N}_0$ is the Euler characteristic of the
entire manifold and $\rho_0(u)$ is the tail probability of the
distribution of the field. (Note that when the manifold is reduced
to a point, this result becomes trivial).

When the level $u$ is high enough, excursions above $u$ become rare
and the excursion set becomes a few disconnected hyper-ellipses. In
that case the Euler characteristic $\phi(A_u)$ simply counts the
number of disconnected components that make up $A_u$. For even
higher levels this number is mostly zero and rarely one, and Its
expectation therefore converges asymptotically to the excursion
probability. We can thus use it as an approximation to the excursion
probability for large enough $u$~\cite{jonatan}

\begin{equation}\label{eq:limit}
\mathbb{E}[\phi(A_u)] \approx \mathbb{P}[ \displaystyle\max_{\theta
\in \mathscr{M}} q(\theta)
> u].
\end{equation}

The practical importance of Eq.~(\ref{eq:euler}) now becomes clear,
as it allows to estimate the excursion probabilities above high
levels. Furthermore, the problem is reduced to finding the constants
$\mathscr{N}_d, d>0$. Since Eq.~(\ref{eq:euler}) holds for any level
$u$, this could be achieved simply by calculating the average of
$\phi(A_u)$ at some low levels, which can be done using a small set
of Monte Carlo simulations. We shall now turn to a specific example
where this procedure is demonstrated.

\section{Application to neutrino source detection}\label{sec3}
\begin{figure}[h!]
\begin{center}
\subfigure[]{\includegraphics[width=7cm]{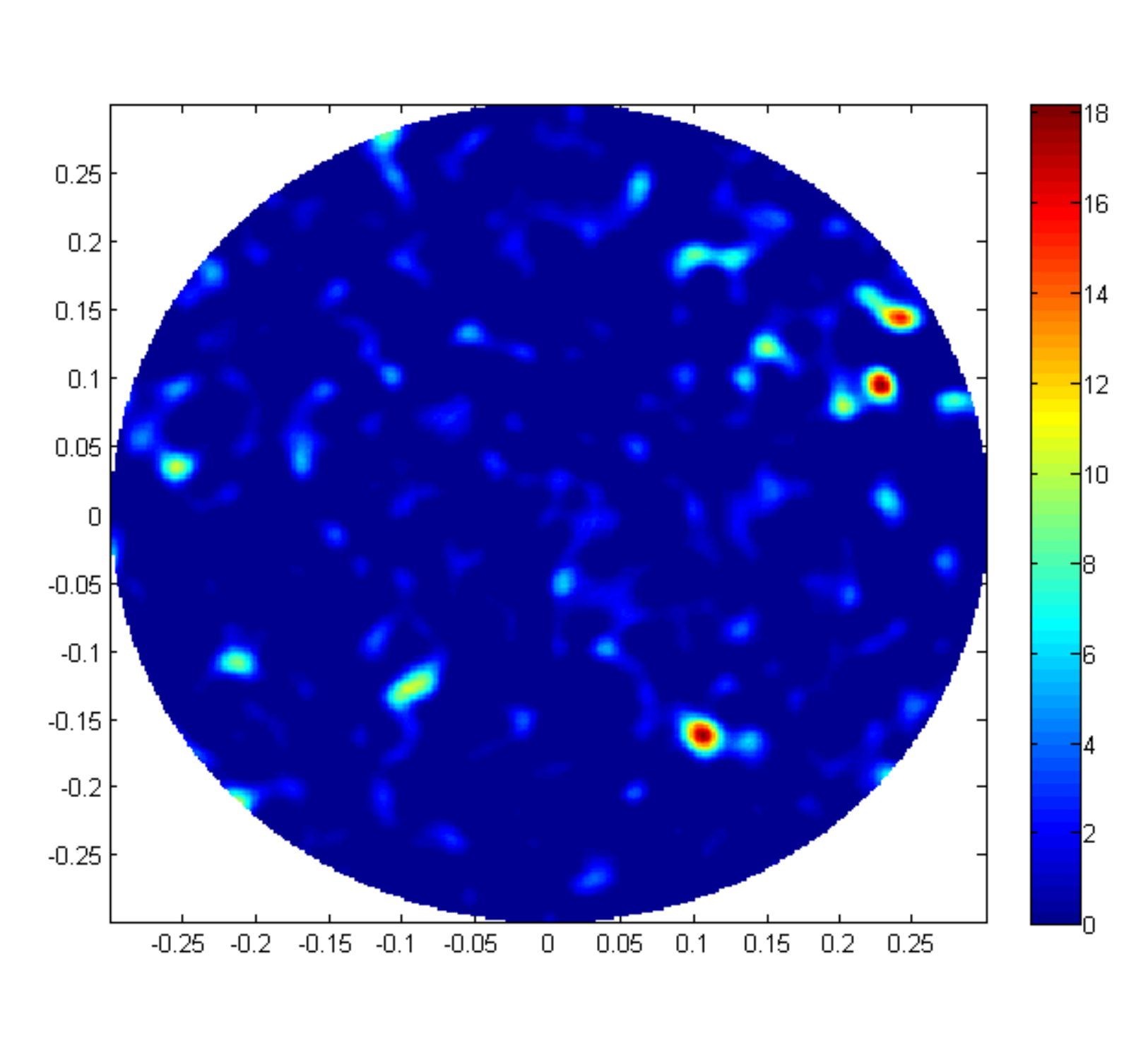}}
\subfigure[]{\includegraphics[width=7cm]{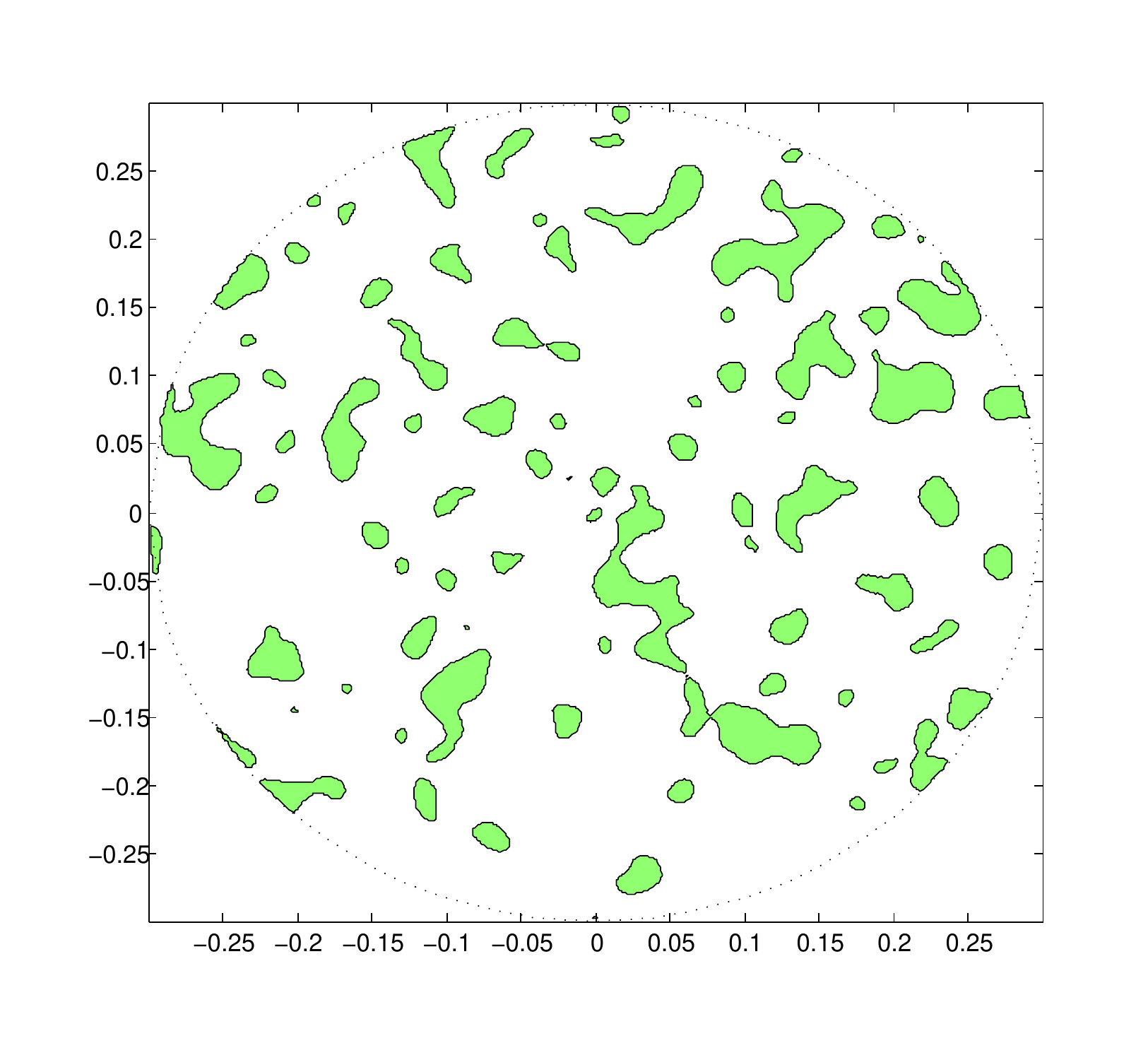}}
\caption{(a) A significance map showing a projection of the test
statistic $q(\theta,\varphi)$ for a background simulation (b) The
corresponding excursion set above $q=1$. The Euler characteristic in
this example is 95.}\label{fig:map}
\end{center}
\end{figure}

The IceCube experiment~\cite{icecube} is a neutrino telescope
located at the south pole and aimed at detecting astrophysical
neutrino sources. The detector measures the energy and angular
direction of incoming neutrinos, trying to distinguish an
astrophysical point-like signal from a large background of
atmospheric neutrinos spread across the sky. The nuisance parameters
over which the search is performed are therefore the angular
coordinates $(\theta,\varphi)$\footnote{The signal model may include
additional parameters such as spectral index and time, which we do
not consider here for simplicity.}. We follow~\cite{teresa} for the
definitions of the signal and background distributions and the
likelihood function. The signal is assumed to be spatially Gaussian
distributed with a width corresponding to the instrumental
resolution of $0.7^o$, and the background from atmospheric neutrinos
is assumed to be uniform in azimuthal angle. We use a background
simulation sample of 67000 events, representing roughly a year of
data, provided to us by the authors of~\cite{teresa}. We then
calculate a profile likelihood ratio as described in the previous
section. Figure~\ref{fig:map} shows a ``significance map'' of the
sky, namely the values of the test statistic $q(\theta,\varphi)$ as
well as the corresponding excursion set above $q=1$. To reduce
computation time we restrict here the search space to the portion of
the sky at declination angle 27$^{\circ}$ below the zenith, however
all the geometrical features of a full sky search are maintained.
Note that the most significance point has a value of the test
statistic above 16, which would correspond to a significance
exceeding 4$\sigma$ if this point would have been analyzed alone,
that is without the ``look elsewhere'' effect.

\subsection{Computation of the Euler characteristic}

\begin{figure}[h!]
\begin{center}
\includegraphics*[width=12cm]{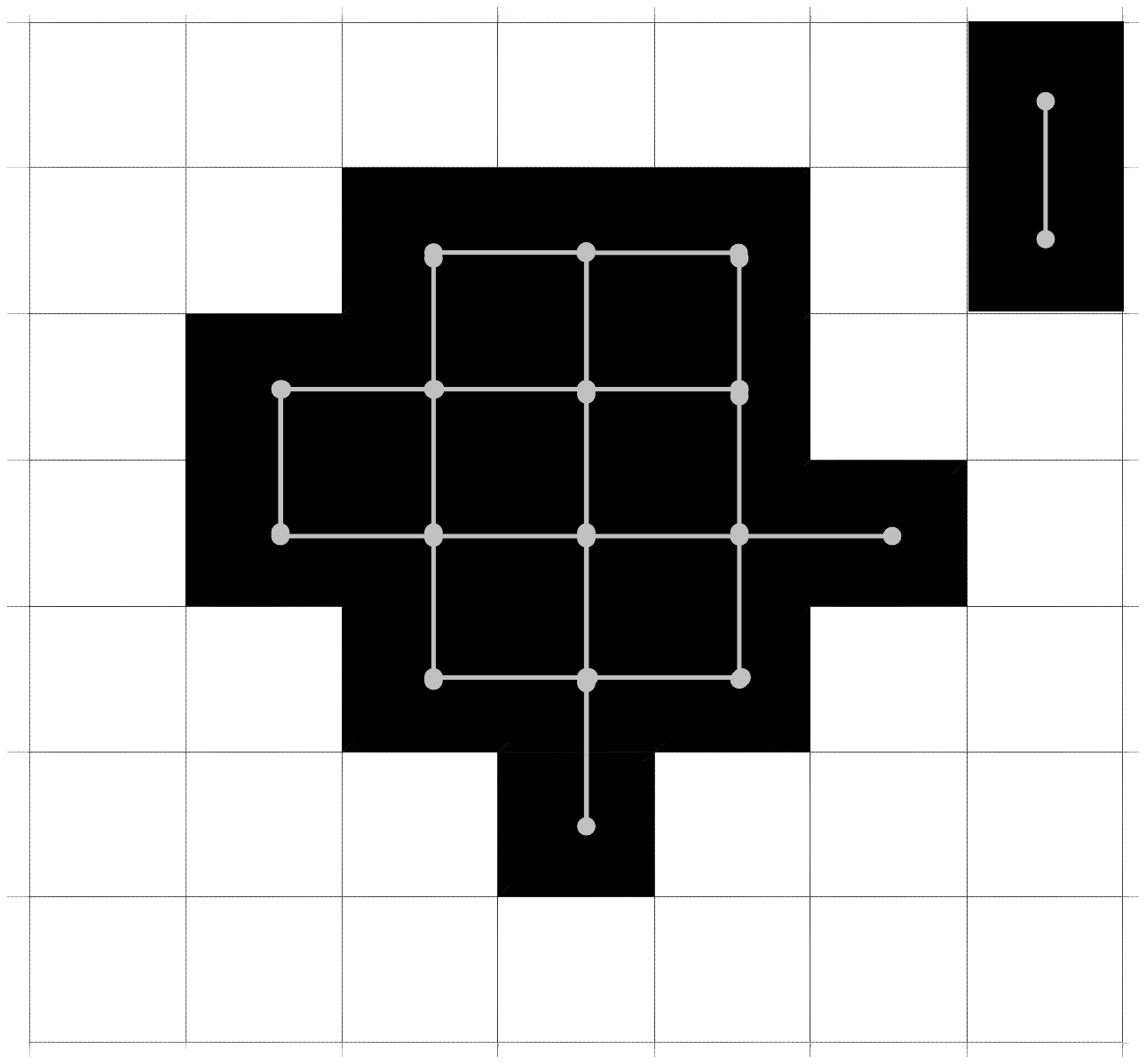}
\end{center}
\caption{Illustration of the computation of the Euler characteristic
using formula~(\ref{eq:ef}). Each square represents a pixel. Here,
the number of vertices is 18, the number of edges is 23 and the
number of faces is 7, giving $\phi=18-23+7=2$. }\label{fig:illus2}
\end{figure}

In practice, the test statistic $q(\theta,\varphi)$ is calculated on
a grid or points, or `pixels', which are sufficiently smaller than
the detector resolution. The computation of the Euler characteristic
can then be done in a straightforward way, using Euler's formula:

\begin{equation}\label{eq:ef}
\phi = V - E + F
\end{equation}

\noindent where $V$, $E$, and $F$ are respectively the numbers of
\emph{vertices} (pixels), \emph{edges} and \emph{faces} making up
the excursion set. An edge is a line connecting two adjacent pixels
and a face is the square made by connecting the edges of four
adjacent pixels. An Illustration is given Fig.\ref{fig:illus2}.
(Although it is most convenient to use a simple square grid, other
grid types can be used if necessary, in which case the faces would
be of other polygonal shapes).

Once the Euler characteristic is calculated, the coefficients of
Eq.~(\ref{eq:euler}) can be readily estimated. For a $\chi^2$ random
field with one degree of freedom and for two search dimensions, the
explicit form of Eq.~(\ref{eq:euler}) is given by~\cite{adler}:

\begin{equation}\label{eq:ex1}
\mathbb{E}[\phi(A_u)] = \mathbb{P}[\chi^2 > u] +
e^{-u/2}(\mathscr{N}_1 + \sqrt{u}\mathscr{N}_2).
\end{equation}

\begin{figure}[ht!]
\begin{center}
\includegraphics[width=8cm,height=7cm]{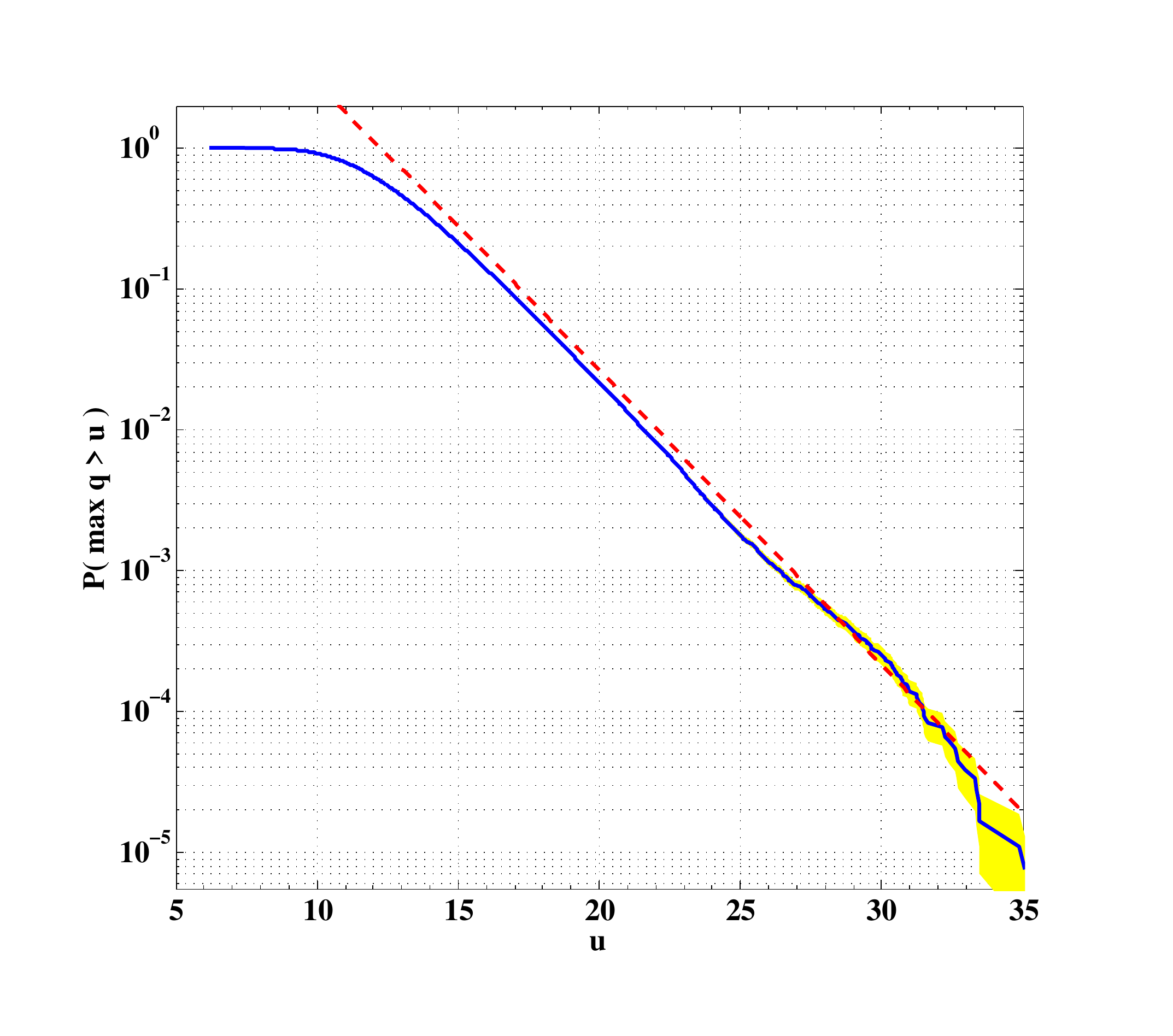}
\end{center}
\caption{The prediction of Eq.~(\ref{eq:ex1}) (dashed red) against
the observed $p$-value (solid blue) from a set of 200,000 background
simulations. The yellow band represents the statistical uncertainty
due to the available number of background simulations.
}\label{fig:2}
\end{figure}

\noindent To estimate the unknown coefficients
$\mathscr{N}_1,\mathscr{N}_2$ we use a set of 20 background
simulations, and calculate the average Euler characteristic of the
excursion set corresponding to the levels $u=0,1$ (The number of
required simulations depends on the desired accuracy level of the
approximation. For most practical purposes, estimating the $p$-value
with a relative uncertainty of about 10\% should be satisfactory.).
This gives the estimates $\mathbb{E}[\phi(A_0)]=33.5 \pm 2$ and
$\mathbb{E}[\phi(A_1)]=94.6 \pm 1.3$. By solving for the unknown
coefficients we obtain $\mathscr{N}_1 = 33 \pm 2$ and $\mathscr{N}_2
= 123 \pm 3$. The prediction of Eq.~(\ref{eq:ex1}) is then compared
against a set of approx. 200,000 background simulations, where for
each one the maximum of $q(\theta,\varphi)$ is found by scanning the
entire grid. The results are shown in Figure~\ref{fig:2}. As
expected, the approximation becomes better as the $p$-value becomes
smaller. The agreement between Eq.~(\ref{eq:ex1}) and the observed
$p$-value is maintained up to the smallest $p$-value that the
available statistics allows us to estimate.

\subsection{Slicing the parameter space}

A useful property of Eq.~(\ref{eq:euler}) that can be illustrated by
this example, is the ability to consider only a small `slice' of the
parameter space from which the expected Euler characteristic (and
hence $p$-value) of the entire space can be estimated, if a symmetry
is present in the problem. This can be done using the
`inclusion-exclusion' property of the Euler characteristic:

\begin{equation}\label{eq:slicing}
\phi(A \cup B) = \phi(A) + \phi(B) - \phi(A\cap B).
\end{equation}

\begin{figure}[h!]
\begin{center}
\includegraphics*[width=8cm]{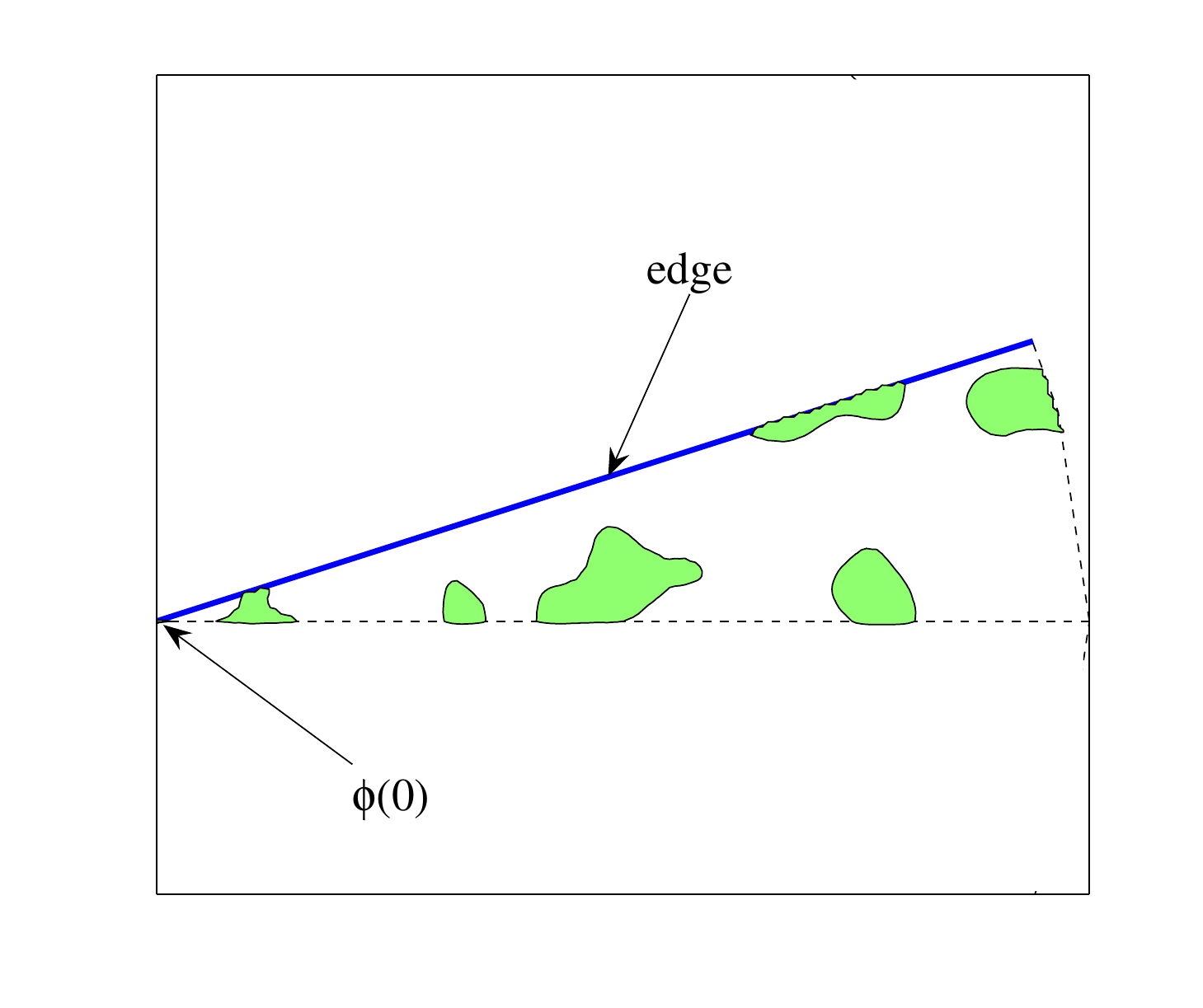}
\end{center}
\caption{Illustration of the excursion set in a slice of a sky,
showing also an edge (solid blue) and the origin as defined in
Eq.~(\ref{eq:euler_slice}). In this example $\phi(slice)=6$,
$\phi(edge)=2$ and $\phi(0)=0$.}\label{fig:slice}
\end{figure}

Since the neutrino background distribution is assumed to be uniform
in azimuthal angle ($\varphi$), we can divide the sky to $N$
identical slices of azimuthal angle, as illustrated in
Figure~\ref{fig:slice}. Applying~(\ref{eq:slicing}) to this case,
the expected Euler characteristic is given by

\begin{equation}\label{eq:euler_slice}
\mathbb{E}[\phi(A_u)] =
N\times(\mathbb{E}[\phi(slice)]-\mathbb{E}[\phi(edge)]) +
\mathbb{E}[\phi(0)]
\end{equation}

\noindent where an `edge' is the line common to two adjacent slices,
and $\phi(0)$ is the Euler characteristic of the point at the origin
(see Figure~\ref{fig:slice}).

\noindent We can now apply Eq.~(\ref{eq:euler}) to both
$\phi(slice)$ and $\phi(edge)$ and estimate the corresponding
coefficients as was done before, using only simulations of a single
slice of the sky. Following this procedure we obtain for this
example with $N=18$ slices from 40 background simulations,
$\mathscr{N}_1^{slice} = 6\pm0.5 , \mathscr{N}_2^{slice} =
6.7\pm0.8$ and $\mathscr{N}_1^{edge} = 4.4\pm0.2$.
Using~(\ref{eq:euler_slice}) this leads to the full sky coefficients
$\mathscr{N}_1=28\pm9$ and $\mathscr{N}_2=120\pm14$, a result which
is consistent with the full sky simulation procedure. This
demonstrates that the $p$-value can be accurately estimated by only
simulating a small portion of the search space.

\section{Summary}
The Euler characteristic formula, a fundamental result from the
theory of random fields, provides a practical mean of estimating a
$p$-value while taking into account the ``look elsewhere effect''.
This result might be particularly useful for experiments that
involve a search for signal over a large parameter space, such as
high energy neutrino telescopes.  While the example considered here
deals with a search in a 2-dimensional space, the formalism is
general and could be in principle applied to any number of search
dimensions. For example, if one is trying to detect a `burst' event
then time would constitute an additional search dimension. In such
case the method of slicing could be useful as well, as one will not
have to simulate the entire operating period of the detector but
only a small 'slice' of time (provided that the background does not
vary in time). Thus, the computational burden of having to perform a
very large number of Monte Carlo simulations in order to to estimate
a $p$-value, could be greatly reduced.

\section{Acknowledgments}
We thank Jim Braun and Teresa Montaruli for their help in providing
us the background simulation data of IceCube which was used to
perform this analysis. One of us (E.~G.) is obliged to the Minerva
Gesellschaft for supporting this work.


\bibliographystyle{model1a-num-names}

\end{document}